\documentclass[11pt]{article}
\usepackage{color}

\usepackage{latexsym}  
\usepackage{amssymb}
\usepackage{graphicx}
\usepackage{amsmath}
\usepackage{mathrsfs}

\newcommand{\be}{\begin{equation}}
\newcommand{\ee}{\end{equation}}
\newcommand{\bea}{\begin{eqnarray}}
\newcommand{\eea}{\end{eqnarray}}
\newcommand{\bref}[1]{(\ref{#1})}

\newcommand{\pa}{\partial}

\begin{document}
\begin{titlepage}
\begin{flushright}
\today
\end{flushright}

\begin{center}
{\Large\bf  
Axion and the Supermassive Black Holes at High $z$}
\end{center}

\begin{center}

\vspace{0.1cm}

{\large Takeshi Fukuyama$^{a,}$%
\footnote{E-mail: fukuyama@rcnp.osaka-u.ac.jp}}

\vspace{0.2cm}

{\small \it ${}^a$Research Center for Nuclear Physics (RCNP),
Osaka University, \\Ibaraki, Osaka, 567-0047, Japan}


\end{center}

\begin{abstract}
Axion dark matter (DM) is studied on the formation of supermassive black holes (SMBHs) at high red shift $z$. It is shown that the attractive self interaction of this DM may resolve the tension between the large mass with high angular momentum of SMBH and its early time formation. We consider the ultra-light axion DM around $O(10^{-21})$ eV as its origin, which may also cause nano-Hz stochastic gravitational wave background recently observed.
\end{abstract}
\end{titlepage}

\section{Introduction}
It is well known that supermassive black holes (SMBHs) ($M_{SMBH}\ge O(10^6M_\odot)$ with solar mass $M_{\odot}$) reside at the center of most galaxies \cite{Richstone1998, Kormendy, Banados}\footnote{Also the supermassive galaxies at $z\geq 10$ is observed by James Webb Space Telescope \cite{JWST}.}. The accretion processes of both baryonic and collisionless DM are rather slow.
So there appears a problem why the SMBHs appear at  high redshift $z\ge 5$. Furthermore, BHs have in general large angular momenta, which makes it more difficult to form BHs at such high $z$. In this paper, we consider axion DM \cite{PQ} and show that its attractive self-interaction may solve the above mentioned problem. 
QCD axion appears in connection with  the chiral condensate, which is the main origin of baryonic masses, and it is natural to consider that axion is the main constituent of DM. 
Dine-Fischler-Srednicki-Zhitnitsky (DFSZ) axion \cite{DFSZ, DFS} is naturally formulated in the framework of the minimal SO(10) GUT \cite{Fuku}. Indeed, in the minimal SO(10) model, there are two Higgs doublets, $H_u$ and $H_d$, coming from ${\bf 10}$-plet and $\overline{126}$-plet Higgs in GUT. They are coupled with ${\bf 210}$-plet Higgs \cite{Fuku2, Babu}, which induces the $H_u-H_d-\phi\mbox{(axion)}^2$ coupling in the Standard Model (SM) phase. The fundamental spinor representation is ${\bf 16}$-plet, including just all SM matters of quarks and leptons, and no room for the additional heavy fermions unlike Kim-Shifman-Vainshtein-Zakharov (KSVZ) axion \cite{KSVZ, SVZ}.  Thus, the DFSZ axion model can be arranged with the minimal SO(10) model by adjusting global $U(1)_{PQ}$ charges. That is, ${\bf 16},~{\bf 10},~{\overline{\bf 126}},~{\bf 210}$ fields have $U(1)_{PQ}$ charge $-1,~+2,~+2,~0$, respectively. As will be shown, string-inspired axion may play the essential role in the formation of the SMBH and it is very suggestive that the axion decay constant $f_a$ is of $O(10^{16})$GeV, coincident with GUT scale (See \bref{GUT}). It is also shown that axion (-like) mass is related with the unified strong coupling (\bref{MSUSY}). However, DFSZ model involves the domain wall problem since $N_{DW}=2N_g=6 (N_g\mbox{:generation number}$) \cite{Babu, Sikivie, Beyer}, whereas KSVZ has $N_{DW}=1$ and no problem. Fortunately enough, axion field acts as $\phi=const.$ in the $m_at\ll 1~(m_a\mbox{:axion mass})$ era and induces the inflation, which dilutes away such topological defect as is shown below \cite{Marsh2016, Soda2020}.  In this letter, we show that the ultra-light axion induces the rapid formation of SMBH and may also be the origin of the stochastic gravitational wave backgroud (SGWB) found by
the North American Nanoherz Observatory for Gravitational Waves (NANOGrav) \cite{Nano1, Nano2} and Pulsar Timing Array Collaborations (PTAs) \cite{Antoniadis, Reardon}.

This letter is organized as follows. In section 2, we review the axion model in the expanding universe, where axion is divided into fast and slow oscillating parts. The slow oscillating part forms overdense part, whose evolution is discussed in section 3. In section 4, we discuss about the problem on the axion model before or during inflation and about SGWB.
We use $\hbar=c=1$ units in this letter.

\section{The review of axion cosmology}
In this section, we review the background of the scenario \cite{Soda2020}.
We introduce the Peccei-Quinn scalar field $\phi$. The axion field $\varphi$ is introduced as its phase
\be
\phi=|\phi|e^{i\theta}= |\phi|e^{i\varphi/f_a}
\label{axion}
\ee
$\phi$ in the expanding universe is described as 
\be
\frac{1}{\sqrt{-g}}\frac{\pa}{\pa x^\mu}\left(\sqrt{-g}g^{\mu\nu}\frac{\pa\phi}{\pa x^\nu}\right)+m^2\phi=0.
\ee
\label{Eq1}

The metric $g_{\mu\nu}$ is
\be
ds^2=(1+2\Phi)dt^2-a^2(1-2\Phi)(dx^2+dy^2+dz^2).
\label{metric}
\ee
Here $\Phi$ is the gravitational potential whose meaning will be discussed later and $a$ is the Friedmann-Robartson-Walker (FRW) metric. We have neglected the interaction with gauge bosons and axion self-coupling since we are concerned with the global behaviours of axion in the expanding universe here. Substituting \bref{metric} into \bref{Eq1}, we obtain
\be
\ddot{\phi}+3H\dot{\phi}-4\dot{\Phi}\dot{\phi}-\frac{1+4\Phi}{a^2}\Delta\phi+m_a^2(1+2\Phi)\phi=0,
\label{phieq}
\ee
where $\dot{\phi}\equiv \frac{\pa\phi}{\pa t}$ and $H=\frac{\dot{a}}{a}$.
Firstly we want to consider the global behaviours of axion and make an approximation
\be
\ddot{\phi}+3H\dot{\phi}+m_a^2\phi=0
\label{Eq2}
\ee
with $H=3p/t$ when $a\propto t^p$. The general solution of \bref{Eq2} is
\be
\phi=a(t)^{-3/2}(m_at)^{1/2}\left(AJ_n(m_at)+BY_n(m_at)\right),
\label{Eq3}
\ee
where $J_n$ and $Y_n$ are Bessel and Neumann functions, respectively, and $n=\frac{1}{2}\sqrt{9p^2-6p+1}$.
Imposing the regularity at $m_at\to 0$, we obtain the integral constant $B=0$.

When $m_at\ll 1$,
\be
J_n(m_at)\approx \frac{1}{\Gamma(n+1)}\left(\frac{m_at}{2}\right)^n,
\label{Eq4}
\ee
and
\be
\phi (t)\propto t^{-\frac{3}{2}p+\frac{1}{2}+n}.
\ee
Then $\phi (t)=const. $ either radiation ($p=\frac{1}{2}$) or matter ($p=\frac{2}{3}$) dominant universe. Thus axion acts as dark energy (DE).

When $m_at\gg 1$, 
\be
J_n(m_at)\approx \left(\frac{2}{\pi m_at}\right)^{1/2}\cos \left( m_at-\frac{n\pi}{2}-\frac{\pi}{4}\right)
\ee
and
\be
\phi(t)=A\sqrt{\frac{2}{\pi}}t^{\frac{3}{2}p} \cos\left( m_at-\frac{n\pi}{2}-\frac{\pi}{4}\right).
\ee
Thus the energy density of axion
\bea
\rho_a&=&\frac{1}{2}\dot{\phi}^2+\frac{1}{2}m_a^2\phi^2\nonumber\\
&=&\frac{A^2m_a^2}{\pi}\frac{1}{a^3(t)}\left[1+\frac{9p^2}{4}\frac{1}{(m_at)^2}\cos ^2\left( m_at-\frac{n\pi}{2}-\frac{\pi}{4}\right)\right] \label{DM1} \\
&\approx & \frac{A^2m_a^2}{\pi}\frac{1}{a^3(t)}.
\eea
Then $\phi$ acts as dark matter (DM) for $m_at\gg 1$

Thus $\phi$ acts as DE for the early $m_at\ll 1$ era and does as DM for the later $m_at\gg 1$ era.
However, we must consider another aspect of axion as a quantum field and the result of Bose Einstein Condensate (BEC).
Aa we mentioned, at $m_at\gg 1$, axion oscillates coherently and indistinguishable with BEC. However, this conclusion comes from \bref{Eq2} with neither gravitational nor axion self interactions. Indeed, the BEC proceeds in a Bose gas of mass $m$ and number density $n$, when
the thermal de Broglie wavelength $\lambda_{dB}\equiv\sqrt{2\pi^2/(mkT_m)}$ exceeds the mean interparticle
distance $n^{-1/3}$, and the wavepacket percolates in space \cite{FMT}:
\be
kT_m < \frac{2\pi^2 n^{2/3}}{m},
\ee
where $n$ is the number density, $n\equiv N/V$. On the other hand, cosmic evolution has the same temperature dependence \cite{FMT} since the
matter-dominant universe behaves, in an adiabatic process, as
\be
\rho\propto T_m^{3/2}.
\ee
Hence, if the boson temperature is equal to radiation temperature at $z = 1000$, for
example, we have the critical temperature at present $T_{critical}= 0.0027$ K, since $T_m \propto a^{-2}$
and therefore $T_\gamma/T_m\propto a$ in an adiabatic evolution. Using the present energy density of
the universe $\rho=9.44 \times 10^{-30}$ g cm${}^{-3}$, the BEC takes place provided that the boson mass
satisfies
\be
m < 1.87\text{eV}.
\ee
However, axion has the self attraction which makes BEC unstable and axion oscillates between
BEC and DM gass.  The difference between BEC and DM occurs on length scales smaller than the de Brogile length. So for the usual QCD axion it is too small to consider \cite{SY}. However, it may not be negligible for the string-inspired axion which will be discussed later.

Axion field potential in the classic dilute gas approximation becomes
\be
V(\phi)=m_u\Lambda_{QCD}^3\left[1-\cos\left(\frac{\varphi}{f_a}\right)\right]\equiv \Lambda^4\left[1-\cos\left(\frac{\varphi}{f_a}\right)\right].
\label{potential}
\ee
Using the Gellmann-Oakes-Renner relation \cite{Gellmann}, 
\be
\Lambda_{QCD}^3=\frac{F_\pi^2m_\pi^2}{m_u+m_d} \footnote{This is the QCD axion in the restricted sence.  QCD axion is involved in more extensive sence like string-inspired axion described later in \bref{potential3}.}.
\label{lambda}
\ee
Here $F_\pi=93$ MeV, and $\frac{m_u}{m_d}\approx 0.47$, and we obtain axion mass
\be
m_a=5.7\times 10^{-6}\left(\frac{10^{12}\mbox{GeV}}{f_a}\right) \text{eV}.
\label{ma}
\ee
Then, the axion self coupling constant $\lambda$ in $\frac{\lambda}{4!}\varphi^4$ becomes
\be
\lambda=-0.47\frac{F_\pi^2m_\pi^2}{f_a^4}<0.
\label{selfc}
\ee
Thus, the self coupling is attraction, and not repulsion unlike the usual Gross-Pitaevskii equation \cite{Gross, Pitaevskii}. As is well known, there are wide ranges of axion mass $m_a$ and breaking energy scale $f_a$, or equivalently $\Lambda$ and $f_a$ but potential form of \bref{potential} is not altered.
In the subsequent sections, we will show that this attractive force may solve the above mentioned problem of the SMBH at high $z$ and confine the ambiguities of axion models.

\section{Slow oscillation part}
 Axion field, $\phi$, is divided into fast oscillation ($e^{im_at}$) part and slow one ($\psi$), as
\be
\phi=\frac{1}{\sqrt{2m_a}}\left(\psi e^{-im_at}+\psi^*e^{im_at}\right).
\label{Eq5}
\ee
This is also the process from the relativistic to nonrelativistic transition.

Substituting \bref{Eq5} into \bref{phieq}, we obtain
\be
i\dot{\psi}+i\frac{3}{2}H\psi+\frac{1}{2m_a^2}\Delta \psi-m_a\Phi\psi=0.
\label{Eqpsi}
\ee
Here we consider the flucutuation (the overdense region) of the Peccei-Quinn field decoupled from the cosmological expansion and use Gaussian approximation \cite{Gupta2017} with angular momentum, 
\begin{equation}
|\psi\left(t,x\right)|=\frac{1}{\sqrt{2\pi(l+1)!\sigma^3}}\left(\frac{r}{\sigma}\right)^le^{-r^2/(2\sigma^2)}Y_{lm}(\theta, \varphi).
\end{equation}
From \bref{Eqpsi} with axion self coupling \bref{potential}, the axion Lagrangian density becomes
 \be
\mathcal{L}= \frac{i}{2}\left(\psi^*\frac{\pa\psi}{\pa t}-\psi\frac{\pa\psi^*}{\pa t}\right)-\frac{1}{2m_a}\nabla\psi^*\cdot\nabla\psi-\frac{gN}{2}|\psi|^4+N|\psi(x)|^2\int\frac{Gm_a^2}{|{\bf x}-{\bf y}|}|\psi(y)|^2d^3y.
\ee
Here $N$ is the total number of axion particles in the overdense region. Dimensional coupling $g$ is related with the above $\lambda$ by
\be
g=\frac{\lambda}{m_a^2}=\frac{4\pi a_s}{m_a},
\label{g}
\ee
where $|a_s|$ is the scattering length. 
Then we obtain the effective potential,
\be
V_{eff}=\frac{1}{2m_a\sigma^2}-\frac{\sqrt{2}}{3\pi}\frac{GNm_a^2}{\sigma}+\frac{l(l+1)}{2m_a\sigma^2}+\frac{gN}{6\sqrt{2}\pi^{3/2}\sigma^3}.
\label{potential2}
\ee
So far we have discussed axion field as DM gass. However, I mentioned this DM acts soon as BEC. However, BEC is not stable also: For the negative scattering length, the energy of axion DM is given by \cite{LY, LL}
\be
\frac{E}{N}=\frac{2\pi a_s n}{m_a}\left[1+\frac{128}{15}\sqrt{\frac{a_s^3n}{\pi}}\right].
\label{gamma1}
\ee
That is, if there is a self-attraction ($a_s<0$), BEC decays to DM gass  at the rate
\be
\Gamma=\frac{128\sqrt{\pi} |a_s|^{5/2}n^{3/2}}{15m_a}.
\ee
Thus 
\be
\mathcal{O}(\Gamma)=\mathcal{O}(|a_s|^3n)^{3/2}\times \frac{1}{m_aa_s^2}.
\label{Gamma3}
\ee
So the BEC half-life time $1/\Gamma$ is very short in comparison with the cosmological time scale over the possible wide ranges of $|a_s|^3n<1$, and DM and BEC are indistinguishable. This conclusion is analogous to that by Sikivie-Yang \cite{SY} except their $a_s^3n$ in place of our $(a_s^3n)^{3/2}$ in \bref{Gamma3}. See also \cite{Guth}. 

Thus we have obtained the main features of axion DM and DE, and let us
proceed to the main problem of Black Hole formation of axion DM taking its self-coupling into consideration. The effective potential \bref{potential2} has the stable orbits at \cite{Eby, Hertzberg}
\be
\sigma_{min}=\frac{\frac{\tilde{l}^2}{2m_a}\pm \sqrt{\left(\frac{\tilde{l}^2}{2m_a}\right)^2+\frac{gGN^2m_a^2}{6\pi^{5/2}}}}
{\frac{\sqrt{2}}{3\pi}GNm_a^2} 
\label{min}
\ee
with 
\be
\tilde{l}^2\equiv l(l+1)+1.
\ee
Here it is very important that axion DM attracts to each others since $g<0$, which allows the two extrema in \bref{min}. The minus (plus) case corresponds to stable (unstable) orbit. These maximum and minimum points coalesce at $\sqrt{{...}}=0$ in \bref{min} and the stable orbit disappears, leading to so called 
Kaup radius,
\be
\sigma_{Kaup}=\frac{\sqrt{3}}{2\pi^{1/4}}\frac{\sqrt{-g}}{\sqrt{G}m_a}=\sqrt{-6\pi^{1/2}g}\frac{M_{Pl}^*}{m_a}
\label{Kaup}
\ee
with the reduced Planck mass $M_{Pl}^*=\sqrt{\frac{1}{8\pi G}}=2.4\times 10^{18}$ Gev and Kaup mass,
\be
M_{Kaup}=Nm_a=\frac{\sqrt{6\pi^{5/2}}}{2}\frac{\tilde{l}^2}{\sqrt{G|\lambda |}}
=25\tilde{l}^2\frac{M_{Pl}^*}{\sqrt{|\lambda|}}.
\label{mass}
\ee
This mass is eventually reduced to BH mass since there is no repulsive force to prevent the collapse,
\be
M_{Kaup}=M_{BH}.
\ee
$\lambda\equiv gm_a^2$ is estimated from \bref{potential} as
\be
\lambda=-\left(\frac{\Lambda}{f_a}\right)^4.
\ee
Then
\be
\lambda=-0.47\frac{(0.093\times 0.140)^2}{10^{48}}=-7.9\times 10^{-54}
\ee
for a dilute gass approximation (\bref{potential}). Substituting this value into \bref{mass},
we obtain $M_{BH}=2.1\tilde{l}^2\times 10^{-10}M_\odot \ll M_\odot$.
If we consider the string-inspired axion \cite{Soda2020, Witten, Arvanitaki, Visinelli}, 
the axion potential roughly becomes
\be
V(\phi)=\Lambda_{string}^4\left[1-\cos\left(\frac{\phi}{f_a}\right)\right]
\label{potential3}
\ee
with
\be
\Lambda_{string}^4=M_{SUSY}^2M_{Pl}^{*2}e^{-S_{instanton}}.
\label{SUSY}
\ee
Here $f_a$ and $m_a$ become independent parameters and are given as follow:  Firstly, $f_a$ is \cite{Witten}
\be
f_a=\frac{\alpha_{GUT}M_{Pl}^*}{\sqrt{2}2\pi}\approx 1.1\times 10^{16}~\mbox{GeV}.
\label{GUT}
\ee
Here we have set the strong coupling constant $\alpha_{GUT}$ as
\be
\alpha_{GUT}=\frac{1}{25}.
\ee
The string-inspired axion mass is
\be
m_a=\frac{M_{SUSY}M_{Pl}^*}{f_a}e^{-\frac{S_{instanton}}{2}}=1.2\times 10^{-14}~\mbox{eV}
\label{SUSYmass}
\ee
with supersymmetry breaking scale $M_{SUSY}$. Here we have used
\be
M_{SUSY}=10^9~\mbox{GeV}, ~~\frac{S_{instanton}}{2}=\frac{\pi}{\alpha_{GUT}}\approx 79
\label{MSUSY}
\ee
and the instanton number $N_{instanton}$,
\be
N_{instanton}=\frac{1}{64\pi^2}\int d^4x\epsilon^{\mu\nu\rho\sigma}F_{\mu\nu}^aF_{\rho\sigma}^a=1.
\ee

Thus we obtain $f_a\approx 10^{16}$ GeV, $m_a\approx 10^{-14}$ eV, and $M_{BH}=50\tilde{l}^2M_\odot$.
However, there are many parameters in axiverse \cite{Arvanitaki}, $M_{SUSY},~S_{instanton},~\mbox{misaligment parameter}~\theta_i,$
primordial isocurvature fraction $\alpha$(\bref{alpha}). Visinelli and Vagnozzi obtained 
\be
S_{instanton}=198\pm 28~~ \mbox{and} ~~M_{SUSY}=10^{11} \mbox{GeV}
\ee
by Bayesian parameter inference in light of many cosmological data \cite{Visinelli} \footnote{This SUSY breaking scale seems to be too large in comparison with that obtained from the gauge coupling unification \cite{Fuku}. However, there are many ambiguities on the intermediate states appearing from GUT to SM scales.}.
Then, if we adopt the center value $198$ for $S_{instanton}$, we obtain
\be
\mbox{log}_{10}(m_a/\mbox{eV})=-21.5^{+1.3}_{-2.3}
\ee
and 
\be
M_{BH}=2.5~\tilde{l}^2\times 10^8M_\odot.
\label{SMBH}
\ee
This may explain the origin of SMBH \cite{Richstone1998, Kormendy, Banados}.
It is very interesting that these values lead to the observed magnitude of DM
\be
\Omega_a\equiv \frac{\rho_a}{\rho_{critical}}=\frac{1}{3H_0^2M_{Pl}^{*2}}\rho_a=\frac{1}{3H_0^2M_{Pl}^{*2}}\times a_{osc}^3\times\frac{1}{2}m_a^2f_a^2
=\sqrt{\frac{m_a}{10^{-27}\text{eV}}}\left(\frac{f_a}{M_{Pl}^*}\right)^2
\ee
with the present Hubble constant $H_0(=10^{-33}$ eV).
Here in the last equality we have used axion with $m_a=10^{-21}$ eV begins to oscillate at $z=10^7$.
Thus $\Omega_a$ is within the observed value $\Omega_a=O(1)$. See \cite{kitajima} on the effect of the reaction of gauge fields.
\section{Discussion}

We have studied the two scenarios of axion models, QCD axion with $f_a=10^{12}$ GeV and the string-inspired axion with $f_a=10^{16}$ GeV.  $f_a=10^{12}$ GeV and $f_a=10^{16}$ GeV correspond to canonical seesaw (or Pati-Salam) and GUT energy scales, respectively in SO(10) GUT theory.  The early formation of SMBH prefers the latter scenario, which may reduce the ambiguities of axiverse parameters.
In either model, axion appears before or during inflation, and is free from the domain wall problem \cite{Sikivie} but induces the isocurvature perturbation \cite{Lyth}.  The latter is constrained by the cosmic microwave background (CMB) \cite{WMAP}.
Axion acquires fluctuations during inflation,
\be
\delta \varphi=f_a\delta \theta=\frac{H_{inf}}{2\pi},
\ee
which leads to isocurvature perturabation $\delta\rho_a$ after it got mass,
\be
\frac{\delta \rho_a}{\rho_a}\approx \frac{H_{inf}}{f_a}.
\ee
and the observed perturbation is constrained as
\be
\alpha\equiv\frac{(\mbox{isocurvature perturbation})^2}{(\mbox{adiabatic perturbation})^2+(\mbox{isocurvature perturbation})^2}<0.077~(95\%CL).
\label{alpha}
\ee
There is an indication that $\alpha=0.05$ is the best fit to the observed power spectrum $P(k)$ \cite{WMAP7}.
However, there is also an indication that CDM isocurvature perturbation is not preferable  at a statistically significant level \cite{Planck2015},
\be
\alpha<0.038~~(95\%~CL).
\ee
Axion also produces non-Gaussianity in general. However, if $\Omega_a h^2=0.2$, the non-Gaussianity is negligible.
Indeed, it is described as \cite{Turner, Kawasaki}
\be
\Omega_ah^2=10^{-3}\left(\frac{f_a}{10^{10}\mbox{GeV}}\right)^{1.2}\left(\frac{H_{inf}/2\pi}{10^{10}\mbox{GeV}}\right)^2.
\ee
Therefore, $f_a=10^{16}$ GeV and $H_{inf}=O(10^9)$GeV give almost full value of DM and we may neglect non-Gaussian contribution.
Concerning with this, we give a comment on the relation among DM and DE.  As we have said, axion acts as DM at $m_at\gg 1$ and have not explained the present value of $\Omega_\Lambda\approx 0.7$. 
If axion dominates DM and DE, DE may remain as BH if
\be
M_{BH}(a)=M_{BH}(a_i)\left(\frac{a}{a_i}\right)^3.
\label{DMDE}
\ee
Here $M_{BH}(a)$ is the total mass of BHs within the FRW scale factor $a$ \cite{Croker}.
There is a positive observational indication on it \cite{Farrah} though also an alternative result \cite{Lei} and a critical comment on the interpretation of \bref{DMDE} \cite{Mistele}. There is an argument that DE appears as a cosmological constant coming from de-Sitter invariant gauge theory of gravitation \cite{deSitter}.

Finally, we add some comments on the discovery of the SGWB by the NANOGrav and PTAs. The purpose of the present paper was to find the reason of the early formation of SMBHs. However, our model also gives the possibility of causing SGWB. 
Indeed, the metric in the Einstein equation is generalized from \bref{metric} to
\be
ds^2=(1+2\Phi(\bf{x},t))dt^2-(1+2\Psi(\bf{x},t))\delta_{ij}dx^idx^j,
\ee
and $\Phi$ and $\Psi$ are divided into the static and oscillation parts \cite{Rubakov},
\be
\Phi(\bf {x},t)=\Phi_0(\bf {x})+\Phi_c(\bf{x})\cos (\omega t+2\alpha(\bf{x}))+\Phi_s(\bf{x})\sin  (\omega  t+2\alpha(\bf{x}))~~\mbox{etc.}
\ee
with
\be
\Delta\Phi_0=4\pi G\rho_{DM} ~\mbox{and}~\Phi_0=-\Psi_0
\ee
etc.
The oscillation part satisfies 
\be
6\ddot{\Psi}-2\Delta (\Psi+\Phi)=8\pi G\rho_{DM}(1+3\cos 2m_at).
\ee
and from the geodesic equation of photon, we obtain
\be
\omega_{obs}=\omega_0+\omega_0\left(\Phi(x_{obs})-\Phi(x_s)\right),
\ee
where $x_s$ and $x_{obs}$ are the positions of the pulsar and observation, respectively. Thus, we observe the monochromatic oscillation of gravitational wave having the amplitude $A$,
\be
A=2\times 10^{-15}\left(\frac{\rho_{DM}}{0.3\mbox{GeV}/cm^3}\right)\left(\frac{10^{-23}\mbox{eV}}{m_a}\right),
\ee
and the frequency $f$,
\be
f=5\times 10^{-9}~\mbox{Hz}\left(\frac{m_a}{10^{-23}\mbox{eV}}\right).
\ee
This may be within 2 $\sigma$ of observations \cite{Nano1}. We need further measurements \cite{SKA} towards more definitive conclusions.

Thus the ultralight axion required for the early formation of SMBHs is one of candidates of origin of SGWB among inspiralling binary SMBH models and cosmic string etc. \cite{Ellis}.

We have considered axion DM models. One is the QCD axion whose potential is given by \bref{lambda} and another is string-inspired axion given by \bref{potential3}.  It has been shown that the string-insired axion may resove the tension of the early formation of primordial SMBHs and also may be the origin of SGWB recently observed.

\section*{Acknowledgments}
We would express our sincere thanks to Dr. J. Soda for useful discussions and comments.
This work is supported by JSPS KAKENHI Grant Numbers JP22H01237.

\end{document}